\documentstyle[sprocl]{article}

\def\be{\begin{equation}}
\def\ee{\end{equation}}
\def\bea{\begin{eqnarray}}
\def\eea{\end{eqnarray}}
\input epsf.tex
\def\desepsf(#1 width #2){\epsfxsize=#2 \epsfbox{#1}}

\begin{document}

\begin{flushright}
PSU-TH/240\\
February 2001\\
\end{flushright}

\vspace*{0.7 cm}

\title{A GAUGE-INVARIANT SUBTRACTION TECHNIQUE FOR
 NON-INCLUSIVE  OBSERVABLES IN QCD\footnote{Talk at the
International Symposium on Multiparticle  Dynamics 
  ISMD2000,   Tihany, Hungary, 9-15 October 2000.}}

\author{F. HAUTMANN}

\address{Department of Physics, Pennsylvania State University,
University Park, PA 16802\\
}


\maketitle\abstracts{Using the electromagnetic form factor of a 
quark as a working example, we describe a subtraction technique 
to treat infrared sensitive regions in non-inclusive processes.}

   Inclusive hard-scattering processes,
 characterized by a single large mass scale, are investigated in QCD
by using asymptotic
freedom and factorization theorems~\cite{eswbook}.
 But the application of QCD to
% less  inclusive, multiple-scale processes,
%which is necessary to study
the study of
multiparticle
final states, 
involving  several   mass scales,
 is much  subtler.
%Factorization theorems for these cases, when they exist,
%are of a more complicated kind.
%
The main practical tool
%to investigate  multiparticle final states
is provided by Monte Carlo event generators, modeling parton shower
and hadronization. In these event generators the theory does not yet
go systematically beyond the leading logarithms. To incorporate
next-to-leading order QCD corrections in parton showers,
extensions of the factorization theorems are necessary, for
which new more precise methods are needed.

An important step in this program is to show how to decompose Feynman 
graphs into terms associated with particular regions in loop  
momentum space. In the case of a 
 Monte Carlo event generator 
simulating the exclusive structure of the hadronic final states,
the observables being computed  are not
 infrared and collinear safe.
It is important to develop techniques~\cite{nonlight,jccmonte,mcsept} 
such that even for such observables the integrands to be associated with 
the ultraviolet region are integrable functions 
 --- and can
 in particular  be integrated numerically
 through a Monte Carlo~\cite{friberg}.

Let us consider as an example
the virtual corrections to the electromagnetic form factor of a
quark (Fig.~1). The theory for this process is well known. See for instance 
the
review in Ref.~\cite{jccmue}.
To simplify the calculations while
retaining all the  ingredients that are 
essential for our discussion, 
let us  work in a massive abelian theory with scalar quarks.
We denote the quark mass by $m$ and the gauge boson mass by $m_g$.
We  work in a center-of-mass frame in which the incoming quark momenta
$p_A$ and  $p_B$ are  in the $+z$ and $-z$ directions, with
$ 2 \,  p_A^+ \, p_B^- = Q^2$.
We consider the amplitude
\begin{eqnarray}
\label{vertex}
&& \Gamma  = i g^2 \int {{d^4 k} \over {(2 \, \pi)^4}}
\\
&& \times {{ (2 \, p_A - k) \cdot (2 \, p_B + k) }
\over {  \left( k^2 - m_g^2 + i \, \varepsilon \right) \,
\left[ (p_A - k)^2 - m^2 + i \, \varepsilon \right] \,
\left[ (p_B + k)^2 - m^2 + i \, \varepsilon \right] \,  } } \,
- {\mbox{UV}}
\nonumber
\hspace*{0.1 cm} ,
\end{eqnarray}
where   $ {\mbox{UV}}$   indicates  the
$\overline{\rm MS}$
counterterm for the ultraviolet
divergence.

Standard power counting arguments~\cite{libby} determine the
 regions of momentum space contributing to the leading power
behavior of $\Gamma$~\cite{jccmue,korchrady}:
1) the  soft region, where
 all components of $k^\mu$ are much smaller than $Q$:
$k^\mu \sim \lambda Q$, with $\lambda$ small;
2) the  $p_A$-collinear region: $k^+ \sim Q$, $k_\perp \sim \lambda \, Q$,
$k^- \sim \lambda^2 \, Q$, with
$ 0 \leq k^+ \leq p_A^+$;
3) the $p_B$-collinear region: $k^- \sim Q$, $k_\perp \sim \lambda \, Q$,
$k^+ \sim \lambda^2 \, Q$, with
$ - p_B^- \leq k^- \leq 0$;
4) the hard region, where all components of $k^\mu$ are of order $Q$.

\begin{figure}[htb]
\centerline{\desepsf(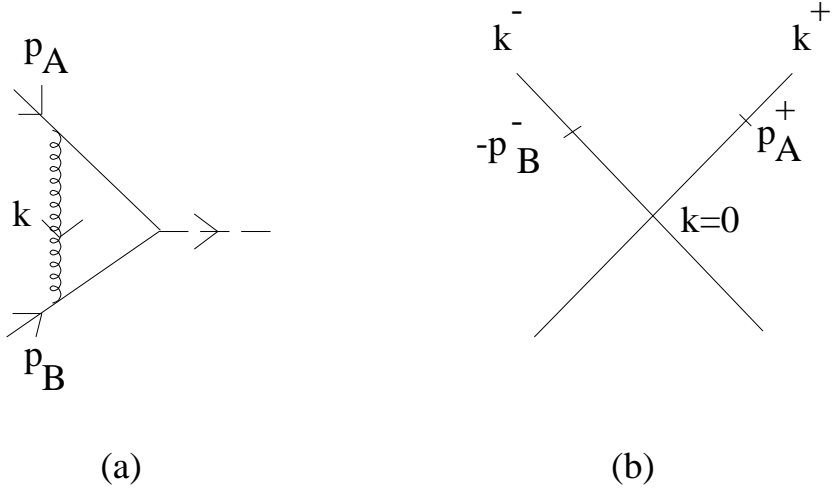 width 11 cm)}
\vspace*{10mm}
\caption{(a) One-loop graph for the electromagnetic
form factor; (b)  geometry of the infrared sensitive regions in
momentum space. }
\label{figformfac}
\end{figure}

Ref.~\cite{nonlight} constructs a decomposition of  $\Gamma$
into a sum of terms, one for each of these regions (Fig.~2),
\begin{equation}
\label{Hterm}
\Gamma = S + A + B + H + {\mbox{nonleading}} \;\; {\mbox{power}} 
\hspace*{0.3 cm} , 
\end{equation}
 satisfying the following requirements:

 i) The splitting between the terms is to be defined gauge-invariantly:
  we demand that  the  terms be obtained from  matrix elements of
 gauge-invariant operators~\cite{jccmue,korchrady}.

ii) In particular, the necessary cut-offs on rapidity integrations should 
  be  gauge-invariant. As we will
 see, this involves the use of Wilson
lines along non-lightlike directions~\cite{nonlight}.

iii) The evolution equations~\cite{eveq,jccmue,korch} 
 with respect to these cut-offs
should be simple, in the sense that there should be no power-law
remainder terms.  All the power-law corrections are associated with the
initial
construction of the terms $S$, $A$, $B$, $H$ in Eq.~(\ref{Hterm}).

iv) The  integrand 
associated with the hard region  should be an integrable function 
even when  the physical observable being computed  is not 
infrared safe in perturbation 
theory.

The strategy we use to construct such a decomposition
is similar to the R-operation techniques for renormalization.
See Ref.~\cite{tka} for a related approach. We proceed from
smaller to larger regions  (Fig.~2).   For each region, we construct a term
that,
added to the terms for smaller regions,
 gives a good  leading-power
approximation to the original amplitude in that
region,  and  does not receive leading contributions from
regions that are smaller or have an overlap with the
region being treated.

To see how this works, let us
look at the form of   the result   \cite{nonlight} for one of  the terms.
The term $S$ associated with the soft region  is
\begin{eqnarray}
\label{Sterm}
S &=&  { {-i \, g^2 } \over {(2 \, \pi)^4}} \,
\int \, d k^+ \, d k^- \, d^2 {\bf k} \, { 1 \over
{ ( k^2 - m_g^2 + i \, \varepsilon ) }}
\, \left[ { 1 \over { ( k^- - i \, \varepsilon ) \,
( k^+ + i \, \varepsilon ) \, } } \right.
\nonumber\\
&&  \hspace*{0.8 cm}
- { 1 \over { ( k^- - i \, \varepsilon )  } } \,
{ { u_B^- } \over { (u_B^- \, k^+ + u_B^+ \, k^- + i \, \varepsilon )}}
\nonumber\\
&&  \hspace*{0.8 cm}
- \left.
{ { u_A^+ } \over { ( u_A^+ \, k^- + u_A^- \, k^+ - i \, \varepsilon )}}
\, { 1 \over { ( k^+ + i \, \varepsilon )  } }
\right]
\nonumber\\
&&
- {\mbox {UV}} \hspace*{0.3 cm}
\end{eqnarray}
The first term in the square brackets
is just obtained by taking the soft approximation to  Eq.~(\ref{vertex}).
This term still has singularities from the
ultraviolet and  collinear regions.  The ultraviolet singularity is 
to be dealt with by the standard subtractive renormalization procedure. 
We treat the collinear singularities 
 in a similar fashion:    
 the next  two terms in the square brackets 
are 
subtractions terms designed to cancel the collinear  contributions. 

\begin{figure}
\centerline{\desepsf(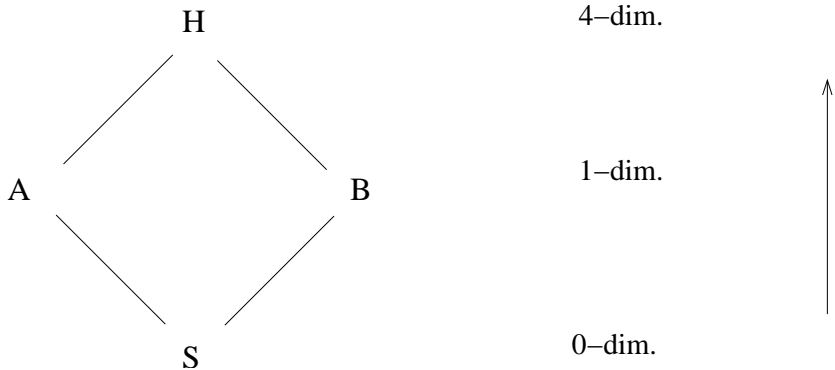 width 11 cm)}
\vspace*{5mm}
\caption{ Soft (S), collinear (A, B) and  hard
(H) contributions.  }
\label{fig:catalog}
\end{figure}

To define these terms  we have  introduced
 two  vectors
$u_A = \left( u_A^+, u_A^-, {\bf 0} \right)$,
$u_B = \left( u_B^+, u_B^-, {\bf 0} \right)$,  lying
along  directions
away from the light cone.
The second term
in the square brackets
subtracts the divergence from the region collinear to
$p_A$, i.e.,
$k^-/k^+ \to 0$.  The non-lightlike vector $u_B$   in
this term provides a cut-off
on the region of  small $k^+$.
Similarly, the third term subtracts the
divergence from the region collinear to $p_B$,  with
the vector $u_A$ providing a cut-off on the region of small  $k^-$.
Further inspection~\cite{nonlight} of the contour 
integrations in Eq.~(\ref{Sterm}) for 
$ k^+ k^- \ll  {\bf k}^2$
 shows that $u_A$ and $u_B$ must be
spacelike,
$u_A^+, u_B^- > 0$, $u_A^-, u_B^+ < 0$.
Note that the 
collinear-to-$p_A$ subtraction term has no collinear-to-$p_B$
singularity; indeed it is power suppressed in this region. The same is
true with $A$ and $B$ exchanged.

The important point is that the cut-offs  thus introduced 
are defined gauge-invariantly: the counterterms in Eq.~(\ref{Sterm}) can
be obtained from matrix elements of  path-ordered exponentials  of the
gluon field along non-lightlike lines.
For a generic direction $n$, define
\begin{eqnarray}
\label{VqVqbar}
V_q ( n ) &=& {\cal P}\exp\left(
 i g \int_{-\infty}^0 dz \, A (z \, n )
 \cdot {n} \right)
\hspace*{0.2 cm} ,  \hspace*{0.3 cm}
\nonumber\\
V_{\bar q} ( n ) &=& {\cal P}\exp\left(
 - i g \int_{-\infty}^0 dz \, A (z \, n )
 \cdot {n} \right)
\hspace*{0.2 cm} .
\end{eqnarray}
Consider  the  product of vacuum expectation values
\begin{equation}
\label{Spath}
{\widetilde S} =
\frac{ 
   \langle 0 | V_q ( {\hat p}_A ) \, V_{\bar q} ( {\hat p}_B ) | 0 \rangle \, 
   \langle 0 | V_q ( {u}_A )  | 0 \rangle \, 
   \langle 0 | V_{\bar q} ( {u}_B )  | 0 \rangle 
}{ 
   \langle 0 | V_q ( {\hat p}_A ) \, V_{\bar q} ( {u}_B ) | 0 \rangle \,
   \langle 0 | V_q ( {u}_A ) \, V_{\bar q} ( {\hat p}_B ) | 0 \rangle 
}
\hspace*{0.2 cm} . 
\end{equation}
At one loop 
the first factor in the
numerator gives the unsubtracted soft term  
in the first line of Eq.~(\ref{Sterm}), while 
 the two factors in the denominator, involving Wilson  lines along
spacelike 
directions, give the collinear subtractions.  The remaining
factors in the numerator cancel factors of a complete external
propagator for the Wilson line.  Given the  one-loop formulas, this
result appears  to be unique, if we simply assume that the
quantity which we calculate is the product of vacuum expectation
values of some Wilson line operators.

The vectors  $u_A$, $u_B$  introduced by the subtractions
are not physical parameters.  Their utility comes from the fact
that evolution equations in $y_A = (1/2) \ln | u_A^+ / u_A^- | $,
$y_B = (1/2) \ln | u_B^+ / u_B^- | $ can be applied to the terms in
Eq.~(\ref{Hterm}) to extract effects associated with large
logarithms~\cite{jccmue,korch}. One of the advantages of the 
subtraction procedure described here is that the 
corresponding evolution equations 
are homogeneous~\cite{nonlight}. This can be contrasted, e.g., with the 
case of Ref.~\cite{jccmue}, where the evolution equations have power-law 
corrections. The simpler structure of the equations may be helpful 
in more complicated cases,  such as 
the    factorization needed  for the  inclusion of 
next-to-leading corrections in 
Monte Carlo event generators.  

The collinear terms $A$, $B$ in Eq.~(\ref{Hterm}) can also be 
given an operator definition in terms of spacelike 
Wilson lines~\cite{nonlight}. The hard term 
$H$ 
is obtained by 
 taking the massless approximation
to $\Gamma-A-B-S$.
The result for $H$ reads~\cite{nonlight} 
\begin{eqnarray}
\label{Hbis}
H &=&    { {-g^2 } \over {8 \, \pi^2}}
\, \int \,{{ d {\bf k}^2 } \over { {\bf k}^2 }}
   \left\{ \ln \left( \frac{ {\bf k}^2 }{ Q^2 } \right)
          +i\pi
          \,+\,
             \frac{1 - {\bf k}^2 / Q^2} {R}
             \left[ \ln \left( \frac{ 1 + R }{ 1 - R } \right)
                    - i \pi
             \right]
   \right\}
 ~-~ {\mbox {UV}}
\nonumber\\
\hspace*{0.3 cm}
\end{eqnarray}
where  $Q^2 =  2 \,  p_A^+ \, p_B^-$ and
\begin{equation}
\label{rootdet}
   R = \left\{
       \begin{array}{ll}
           \sqrt{1- 4 \, {{\bf k}^2 /Q^2}}
           &
           ~ ~ ~\mbox {if $4 {\bf k}^2 /Q^2 \leq 1$}
           \hspace*{0.3 cm} ,
       \\[2mm]
           i \, \sqrt{ 4 \, {{\bf k}^2 /Q^2} - 1}
           &
           ~ ~ ~ \mbox {if $4 {\bf k}^2 /Q^2 >  1$}
           \hspace*{0.3 cm} .
       \end{array}
       \right.
\end{equation}
$H$ is independent of the choice of the vectors $u_A$, $u_B$. 
As a result of the infrared subtractions, the
${\bf k}^2$ integration in Eq.~(\ref{Hbis}) is regular at small
${\bf k}^2$.  The large ${\bf k}^2$ behavior is
to be dealt with via an ultraviolet counterterm. 

In conclusion, the subtractive procedure 
that we have applied 
gives a finite coefficient 
for the hard part of the form factor, while the counterterms have  
 a simple 
meaning in terms of gauge-invariant operators and obey  homogeneous 
evolution equations. 
Similar procedures can be  defined in real emission processes 
for collinear~\cite{jccmonte} and soft~\cite{mcsept} contributions.   
Calculational schemes of this kind 
 will be needed to improve the accuracy of 
Monte Carlo calculations for multiparticle final states 
beyond the leading order.

\section*{Acknowledgments}
The results presented
in this talk have been obtained in collaboration with J.~Collins.
This research is supported in part by the US Department of
Energy. I thank the organizers of ISMD2000 for their invitation
and for the excellent organization of the conference.

\end{document}